\documentclass{aastex}

\shorttitle{NIR spectra of ISOHDFS galaxies} \shortauthors{Rigopoulou
et al.}

\begin{document}


\title{VLT-ISAAC near-IR Spectroscopy of ISO selected Hubble Deep
 Field South Galaxies\footnote{Based on observations
with ISO, an ESA project with instruments funded
by ESA member states (especially the PI countries: France, Germany, the
Netherlands, and the United Kingdom) with the participation of ISAS and NASA.}
\footnote {Based on observations collected at the European Southern
Observatory, Chile, ESO No 63.O-0022}   }


\author{D. Rigopoulou \altaffilmark{1}, A. Franceschini\altaffilmark{2},
H. Aussel\altaffilmark{3}, R. Genzel\altaffilmark{1}, 
P. van der Werf\altaffilmark{4}, C.J. Cesarsky\altaffilmark{5},
M.Dennefeld\altaffilmark{6}, S. Oliver\altaffilmark{7},
M. Rowan-Robinson\altaffilmark{7}, R. Mann\altaffilmark{8},
I. Perez-Fournon\altaffilmark{9}, B. Rocca-Volmerange\altaffilmark{6}}
\altaffiltext{1}{Max-Planck-Institut f\"ur
extraterrestrische Physik,  Postfach 1312, 85741 Garching, Germany}
\altaffiltext{2}{Dipartimento di Astronomia, Universita' di Padova,  
Vicolo Osservatorio 5 I-35122, Padova, Italy}  
\altaffiltext{3}{Osservatorio Astronomico di Padova, Vicolo Osservatorio 5
I-35122, Padova, Italy} 
\altaffiltext{4}{Leiden Observatory, PO Box 9513, 2300 RA, Leiden,  The
Netherlands}
\altaffiltext{5}{European Southern
Observatory, Karl-Schwarzschild-str. 2,  85740 Garching, Germany}
\altaffiltext{6}{Institut d'Astrophysique de Paris -
CNRS, 98 bis Boulevard  Arago, 75014 Paris, France}
\altaffiltext{7}{ICSTM,  Astrophysics Group,
Blackett Laboratory, Prince Consort Rd.,  London, SW2 1BZ, U.K.}
\altaffiltext{7}{Institute for Astronomy, University of Edinburgh, 
Royal Observatory, Blackford Hill, Edinburgh, EH9 3NJ, UK}
\altaffiltext{9}{Instituto de Astrofisica de
Canarias, Via Lactea s$/$n,  38200 La Laguna, Tenerife, Spain}
%

\begin{abstract}

We report the results of near-infrared VLT-ISAAC spectroscopy of a
sample of 12 galaxies at z = 0.4--1.4,  drawn from the ISOCAM survey
of the Hubble Deep Field South.  We find that the rest frame R-band
spectra of the ISOCAM galaxies resemble those of powerful
dust-enshrouded starbursts.  H$_{\alpha}$ emission is detected in 11
out of 12 objects down to a flux limit of 7$\times$10$^{-17}$ erg
cm$^{-2}$ s$^{-1}$, corresponding to a luminosity limit of 10$^{41}$
erg s$^{-1}$ at z = 0.6, (for an H$_{\rm o}$ = 50 and $\Omega$ = 0.3
cosmology). From the H$_{\alpha}$ luminosities in these galaxies we
derive estimates of the star formation  rate in the range 2--50
M$_{\odot}/yr$ for stellar masses  1--100 M$_{\odot}$.  The raw
H$_{\alpha}$--based star formation rates are an order of magnitude or
more lower than SFR(FIR) estimates based on ISOCAM LW3 fluxes.   If
the H$_{\alpha}$ emission is corrected for extinction  the median
offset is reduced to a factor of 3.  The sample galaxies are part of a
new population of optically faint    but infrared--luminous active
starburst galaxies, which are characterized by an extremely high rate
of evolution with redshift up to z$\sim$1.5 and expected to contribute
significantly to the cosmic far-IR extragalactic background.

\end{abstract}


\section{Introduction}

Until recently, most of our knowledge about high-z galaxies has come
from optical surveys.  The COBE detection of an extragalactic
far--infrared$/$submm background  (Puget  et al. 1996), with an
integrated intensity similar to or  greater than that of optical light
(e.g. Hauser et al. 1998), strongly suggests that a significant
fraction of the cosmic star formation in the Universe is  obscured by
dust and thus missed by the various optical surveys.

With the advent of the Infrared Space Observatory (ISO,  Kessler et
al. 1996) deep mid--IR surveys for distant galaxies, have
been successfully carried out for the first time.  
Operating in the 5 -- 18 $\mu$m band
sensitive to warm dust and emission from Polycyclic Aromatic Hydrocarbons 
(PAH), ISOCAM on
board ISO was more than 1000 times more sensitive than IRAS and  thus
had the potential to study infrared bright galaxies at redshifts
beyond 0.5.  A number of cosmological surveys have been performed with
ISOCAM especially in the LW3 filter (12 -- 18$\mu$m). 
These surveys range from the wide and shallow
European Large Area ISO Survey (ELAIS, Oliver et al. 2000), to
deep pencil--beam surveys in the Lockman Hole,  the Marano field,
the Northern and Southern Hubble Deep Fields (HDF--N, HDF--S)
and the distant cluster Abell
2390, reaching limiting flux densities of 50 -- 100 $\mu$Jy  (Elbaz
et al. 1999).

At the bright end the ISOCAM source counts combined with those of IRAS
are in good agreement with no or moderate evolution.  
At fainter flux densities the counts steepen considerably and
at $\sim$ 200 -- 600 $\mu$Jy  they are about an order of
magnitude greater than the predictions of no evolution models. This
steepening in the log N--log S plot and a pronounced maximum in the
differential number counts  at $\sim$ 400 $\mu$Jy suggest that the
ISOCAM surveys have  revealed a population of strongly evolving
galaxies. Elbaz et al. (2000, in preparation) show that this population 
plausibly accounts for a significant fraction of the far-IR background.
The next step is to explore the nature of the ISOCAM population
with optical$/$near-IR spectroscopy. In this letter we report on the 
first near-IR (rest-frame R--band) spectroscopic survey of a representative 
sample of faint ISOCAM galaxies in the HDF--S field.

\section{Sample Selection}

The HDF--S was observed by ISOCAM as  part of the ELAIS survey.  The
observations were carried out at two wavelengths,  LW2 ( 6.75 $\mu$m)
and LW3 (15 $\mu$m). Oliver et al. (2000, in preparation)  and Aussel
et al. (2000, in preparation) analyzed the data independently. Aussel
et al.  used the  PRETI method and detected 63 sources  brighter than
S$_{15 \mu m}$ = 100 $\mu$Jy in the LW3 band.  We used this source
list as input for our observations.

We selected the sample for ISAAC follow up from the HDF--S LW3 sources
based on the following criteria: a) a reliable LW3 detection, b)
H$_{\alpha}$ in the wavelength range of ISAAC and, c) a secure
counterpart in the I band image (Dennefeld et al. 2000, in
preparation), or a counterpart in the K band image (ESO Imaging Survey
(EIS) Deep).    We did not apply any selection based on colors. Our
reference sample contains 25 galaxies with 15 $\mu$m flux densities
ranging between 100--800 $\mu$Jy.  It is thus a fair representation of
the strongly evolving ISOCAM population near the peak of the
differential source counts (Elbaz et al. 1999). From these 25
optically identified sources we randomly selected 12 sources for
ISAAC--follow up.  To select the near-IR band (Z, SZ, J, H) for our
spectroscopy we used spectroscopic redshifts from optical
spectroscopy, where available, for z$<$0.7 (Dennefeld et al. 2000, in
prep.).   Otherwise we used photometric redshift estimates based on
the model PEGASE (Fioc and Rocca-Volmerange 1997). Our photometric
redshift determinations turned out to be accurate to
$\Delta$z$\pm$0.1.

\section{Observations and Results}

We collected the spectra  during 1999 September 20--24 with the
infrared spectrometer ISAAC (Moorwood et al. 1998) on  the ANTU--ESO
telescope (formerly UT1), on Paranal, Chile.  For the observations we
used the low resolution grating R$_{s} \sim $ 600 and a
1$^{\arcsec}$$\times$2$^{\arcmin}$  long slit.  To maximize the
observing efficiency each slit position included on average  two
galaxies at any given orientation.  Most of the targets were first
acquired directly from a 1--2 min exposure in the  H-band. In the case
of the very faint objects (H $\geq$ 20.0 mag) we offset from a
brighter star in the HDF-S field.  Observations were made by nodding
the telescope $\pm$ 20$^{\arcsec}$  along the slit  to facilitate sky
subtraction (always avoiding overlap of the two objects in the
slit). Individual exposures ranged from 2--4 minutes. Sky conditions
were excellent throughout the acquisition of the spectra,  with seeing
values in the range 0.4$^{\arcsec}$--1.0$^{\arcsec}$.  For each
filter, observations of spectroscopic standard stars were made in
order to flux calibrate the galaxy spectra.

The data were reduced using applications from ECLIPSE
(Devillard 1998) and IRAF packages.  Accurate sky subtraction is
critical to the detection of faint lines.  Sky was removed by
subtracting the pairs of offset frames. In some cases this left a
residual signal (due to temporal sky changes) which was then removed
by performing a polynomial interpolation along the slit.  OH sky
emission lines were also carefully removed from the spectra.  Spectrum
extraction for  each galaxy was performed using the APEXTRACT package.
Standard wavelength calibration was applied.

The spectra for all 12 galaxies observed with ISAAC are shown in
Figure 1. Table 1 contains exposure times, H$_{\rm AB}$ magnitudes,
measured spectroscopic redshifts (from H$_{\alpha}$ detections),
H$_{\alpha}$ line fluxes and Equivalent Widths (EW) (and where
resolved [NII] fluxes).  We convert the H$_{\alpha}$ line fluxes to
luminosities using H$_{\rm o}$ = 50 km s$^{-1}$ Mpc$^{-1}$ and
$\Omega$ = 0.3.   We also list in Table 1 FIR luminosities based on
LW3 fluxes (see Section 5 for more details).  We note that we have not
detected any H$_{\alpha}$ Broad Line components.

\section{The nature of the ISOCAM faint galaxies: Dusty and Luminous 
Starbursts}

Prior to our study no near-infrared (rest-frame R-band) spectroscopy
had been carried out for the ISOCAM population, primarily because of
the faintness of the galaxies. Aussel et al. (1999) and Flores et al.
(1999) have presented optical spectroscopic analysis (rest-frame
B-band) for HDF-N and the 1415$+$52 field of the  Canada-France
Redshift Survey (CFRS),  respectively. Aussel et al. (1999)
cross-correlated the ISOCAM HDF-N galaxies with the optical catalog of
Barger et al. (1999) resulting in 38 galaxies with confirmed
spectroscopic redshifts.  Flores et al. have identified 22 galaxies
with confirmed spectroscopic information. In both of these samples the
median redshift is about 0.7. Our ISOHDFS sample contains 
7 galaxies 0.4$<$z$<$0.7 and 5 galaxies with
0.7$<$z$<$1.4 and thus, has a z-distribution very similar to
the HDF-N (Aussel et al.) and CFRS (Flores et al.) samples.

Rest-frame B-band spectra  host a number of emission and absorption
lines related to the properties of the starburst in the galaxy. Based
on these features galaxies can be classified according to their
starburst history. Strong H$_{\delta}$, H$_{\epsilon}$ Balmer
absorption and no emission lines are characteristic of  passively
evolving k$+$A galaxies.  The presence of Balmer absorption lines
implies the presence of a dominating A-star population formed about
0.1--1 Gyr ago.   The simultaneous presence of Balmer absorption and
moderate flux [OII] and H$_{\beta}$ emission termed as e(a) or S$+$A
galaxies, indicates that, in addition, there is ongoing star
formation. The relative importance of these star formation episodes
depends on the extinction (especially of the current star formation
component). If the extinction is low the galaxy is primarily a
post--starburst system. If the extinction toward the star forming
region is high then the galaxy could be a powerful starburst. The
majority of the galaxies in the CFRS field ($\sim$ 70\%) display
optical spectra characteristic of e(a) galaxies.  As evidenced by the
detections in Figure 1, the ISOCAM galaxies are in fact powerful
starbursts hidden by large amounts of dust extinction.

Remarkably, dusty starbursts such as M82  (L$\sim$ 10$^{10}$
L$_{\odot}$, Kennicutt et al. 1992), LIRGs (Luminous InfraRed
Galaxies,  L$\sim$ 10$^{11}$ L$_{\odot}$, Wu et al. 1998), and many
bright ULIRGs (L$\sim$ 10$^{12}$ L$_{\odot}$, Liu and Kennicutt 1995),
show e(a) B--band spectra.  Local e(a) galaxies have large
H$_{\alpha}$ equivalent widths (EW), at the same time demonstrating
active current star formation and differential dust extinction.  The
measured EW ratio ([OII]$/$H$_{\alpha}$) for e(a) galaxies appears to
be somewhat low. Such low ratios have already been observed in the
spectra of distant clusters (Dressler et al. 1999), nearby mergers
(Poggianti and Wu 2000), the dusty LIRGs studied by Wu et al. (1998)
or the interacting$/$merging systems studied by  Liu and Kennicutt
(1995). The behaviour of the (OII)$/$(H$_{\alpha}+$NII) ratio is shown
in the EW(OII)--EW(H$_{\alpha} +$ NII) diagram of Figure 2: the
majority of the points lie below the straight line. For our ISOHDFS
sample we use the EW(H$_{\alpha}$) measured from the  observations
presented here.  For the EW(OII) we use a median value of 20$\pm$15
\.A which was recently measured from FORS2 I-band spectra of a small
sample of ISOHDFS galaxies (Franceschini et al. 2000, in
preparation). This value is in agreement with the results presented by
Flores et al. (1999) for the CFRS galaxies.

It follows from Figure 2 that the ISOHDFS galaxies occupy the same
region in the EW(OII)$/$EW(H$_{\alpha} +$ NII) diagram as actively
starforming galaxies. Intrinsic differential dust extinction is
responsible for the somewhat low EW(OII)$/$ EW(H$_{\alpha}+$NII)
ratio. The [OII] emission is affected more than H$_{\alpha}$ simply
because of  its shorter wavelength.   The continuum is due to A-stars
which come from earlier (0.1--1.0 Gyr) star formation activity that is
not energetically dominant and plays a small role once the dusty
starburst is dereddened.  This scenario implies that these galaxies
undergo multiple burst events: the less extincted population is due to
an older burst while in the heavily dust enshrouded HII regions there
is ongoing star formation.  We conclude that ISOCAM galaxies are
actively starforming,  dust enshrouded galaxies, akin to local LIRGs
(e.g. NGC 3256, Rigopoulou et al. 1996).

\section{Star Formation Rates and Extinction Corrections}

The conversion factor between ionizing luminosity and star formation
rate (SFR) is usually computed using an evolutionary synthesis
model. Only massive stars ($>$ 20 M$_{\odot}$) with short lifetimes
($\leq$ 10$^{6}$ yrs) contribute to the integrated ionizing flux.
Using the stellar synthesis code STARS  (Sternberg 1998) we create
models for solar abundances, a Salpeter IMF (1--100 M$_{\odot}$)  and
slowly decaying bursts with ages in the range of a few$\times$
10$^{7}$--10$^{8}$ yrs, and SFR decay time-scales in the range
10$^{7}$-10$^{9}$ yrs. Averaging, we obtain:
\begin{equation}
$$
SFR(M$_{\odot}/yr$) = 5 $\times$10$^{-42}$L$_{H_{\alpha}}$(erg s$^{-1}$).
$$
\end{equation}
We have used this formula to estimate the SFR rates  in Table 2. The
SFR estimates based on Eqn.(1) are a factor of 1.6 smaller than the
SFR estimates based on the Kennicutt (1998) relationship that refers
to stars in the range  0.1--100 M$_{\odot}$.  Averaging over our
models, the SFR scales with the FIR luminosity as:
\begin{equation}
$$
SFR(M$_{\odot}/yr$) = 2.6 $\times$10$^{-44}$L$_{FIR}$(erg s$^{-1}$)
$$
\end{equation}
Since extinction is at play, the SFR estimates we quote in the first
column of Table 2 are {\em lower limits} to the real SFR in these
galaxies.  We derive the extinction based on V--K color indices
(magnitudes taken from the  EIS Survey). Using STARS  as well as the
Starburst99 (Leitherer et al. 1999) codes for various star formation
histories (ie bursts of different duration, and continuous star
formation) we calculate the range of intrinsic colors. The model
predicted intrinsic V--K colors are in the range 1.1 -- 1.5. We have
applied infrared and optical K-corrections from Poggianti (1997) and
Coleman (1980), respectively.  Comparing the observed V--K colors to
the predicted ones we obtain a median  color excess of 2.0 which
corresponds to a median  A$_{\small V}$ of 1.8 assuming a screen model
for the extinction.  This A$_{\small V}$ value corresponds to a median
correction factor for the SFR(H$_{\alpha}$) of $\sim$4.

The SFR can also be inferred from far-infrared (FIR) luminosities
according to Eqn. (2).  The SFR(FIR) estimates in Table 2 are based on
the method of Franceschini et al. (2000, in prep.) which makes use of
the 15 $\mu$m flux and assumes a L$_{FIR} /$L$_{MIR}$ ratio of
$\sim$10 (for an M82 like SED, Vigroux et al. 1999).   The SFR(FIR)
estimates turn out to be a factor of 5 to 50 higher than the SFR
estimates inferred from the {\em non-extinction corrected}
H$_{\alpha}$.  However, if we apply the correction factor of $\sim$4
we deduced for the H$_{\alpha}$ then
SFR(FIR)$/$SFR(H$_{\alpha}$)$\sim$ 3, confirming that the extinction
is much higher than can be predicted using  (UV or) optical
observations.  Thus, ISOCAM galaxies are in fact actively star forming
highly dust enshrouded galaxies. One exception is source ISOHDFS 38
for which we have evidence for the presence of a dominant AGN
component (both from the LW2$/$LW3 ratio and the H$_{\alpha}/$[NII]
ratio).

Based on our SFR estimates, and the evidence for exinction 
presented in section 4.1 we conclude that ISOCAM galaxies are indeed 
dust enshrouded actively star-forming galaxies and not decaying 
post-starburst systems.

\section{Conclusions}

We have presented NIR spectroscopy, rest frame R-band, of a sample of
ISO selected galaxies from the Hubble Deep Field South. We have
detected H$_{\alpha}$ emission in almost all of them. The detections
of the H$_{\alpha}$ line  combined with the large H$_{\alpha}$ EWs
are consistent with the idea that these galaxies are ongoing 
powerful dusty starbursts.

Using the observed H$_{\alpha}$ emission lines we estimate that the
SFR rate in the ISOHDFS galaxies ranges between 2 and 50 M$_{\odot} /$
yr, far higher than those inferred from local Starbursts  (Calzetti
1997) and local spirals (Kennicutt 1992).  We have compared these
rates of SF with the values estimated from the  FIR luminosities,
which are typically a factor 5 to 50 larger because of dust
obscuration. We estimated the H$_{\alpha}$ extinction  using standard
extinction laws.  The H$_{\alpha}$ extinction corrected SFR estimates
are then higher although still fall short of the SFR estimates based
on FIR luminosities ( SFR(FIR):50--400 M$_{\odot}/$ yr).  This result
demonstrates that it is very dangerous to derive star formation rates
from UV or optical data alone since these wavelengths are susceptible
to higher extinction. Thus a significant fraction of star formation is
missed by optical surveys.  We conclude that ISO has detected in the
mid-IR the most active, luminous and dust-enshrouded starbursts at
z$\sim$ 0.4-1.4, which would have remained  otherwise unnoticed by
optical surveys.  This population of strongly evolving active dusty
starbursts  is likely to account for a substantial fraction of the
FIR$/$submm background (Elbaz et al. 2000, in preparation).

\acknowledgments This work is supported by the EC TMR Network
``European Large Area ISO Surveys'' (contract No.
ERBFMRX-CT96-0068). We thank the EC TMR Network ``Galaxy Formation and
Evolution '' for making  redshift information available prior to
publication.  H.A. is supported by the  TMR network ``Galaxy formation
and evolution'', contract  No. ERBFRX-CT96-0086. We thank Amiel
Sternberg for fruitful discussions.

\newpage

\centerline{\bf FIGURES}

\figcaption{ISOHDFS-VLT spectra}

\figcaption{EW(OII) -- EW(H$_{\alpha} +$NII)}


\begin{deluxetable}{c c c c c c c c}
\tablecolumns{8} 
\tablewidth{0pc} 
\tablenum{1} 
\tablecaption{ISOHDFS sample and flux measurements} 
\tablehead{
\colhead{name}& \colhead{Exp.}&
\colhead{z$_{spec}$}&\colhead{Hmag}&
\colhead{F(H$_{\alpha}$)$^{1}$}&\colhead{EW(H$_{\alpha}+$[NII])}& 
\colhead{L(H$_{\alpha}$)}&\colhead{L(FIR)}\\
\colhead{}&\colhead{}& \colhead{}&\colhead{}&\colhead{F([NII])$^{1}$}&
\colhead{EW([NII])}&\colhead{}&\colhead{}\\
\colhead{}&
\colhead{(sec)}& \colhead{}& \colhead{(mag)}& 
\colhead{}& \colhead{\.A}&  \colhead{}&
\colhead{}}
\startdata 
ISOHDFS16&3720&0.62&20.73&1.17&45&0.29&0.54  \\
ISOHDFS23&3720&0.46&19.26&1.86&50&0.23&2.5 \\ 
ISOHDFS25&3720&0.59&19.80&3.12&110$^{5}$&0.67&2.0  \\ 
 & & & &0.72&35& \\
ISOHDFS27&3720&0.58&18.39&3.28&47&0.71&1.61  \\
ISOHDFS28&3720&0.56&20.36&0.78&47&0.16&0.77\\
ISOHDFS38&7400&1.39&21.67&1.95&35$^{5}$&3.62&23.0  \\     
 & & & &4.22& 50& \\
ISOHDFS39&7400&1.27&20.95&7.13&67&10.4&6.15  \\ 
ISOHDFS43$^{4}$&3720&--&21.92&--&--&--&--\\
ISOHDFS53&3720&0.58&19.47&6.08&70$^{5}$&1.32&1.53\\
 & & & &2.8&50& \\
ISOHDFS55&3720&0.76&20.05&2.41&40&0.98&1.27  \\
ISOHDFS60&7400&1.23&20.63&2.73&44&3.69&2.42  \\
ISOHDFS62&3720&0.73&20.67&2.54&62&0.95&1.19  \\
\enddata
\tablenotetext{1}{in units 10$^{-16}$ erg cm$^{-2}$s$^{-1}$}
\tablenotetext{2}{L(H$_{\alpha}$) in units 10$^{42}$ ergs$^{-1}$}
\tablenotetext{3}{L(FIR) in units 10$^{45}$ ergs$^{-1}$}
\tablenotetext{4}{no emission line detected}
\tablenotetext{5}{H$_{\alpha}+$[NII] emission resolved. 
H$_{\alpha}$ and [NII] fluxes and EWs are reported separately.}
\end{deluxetable}

\begin{deluxetable}{c c c c}
\tablecolumns{4} 
\tablewidth{0pc} 
\tablenum{2}
\tablecaption{Star Formation Rates} 
\tablehead{ \colhead{name}&
\colhead{SFR(H$_{\alpha}$)}& \colhead{SFR(FIR)}\\
\colhead{}&\colhead{\small{uncorrected for extinction}}&
\colhead{\small{based on FIR estimates}} } 
\startdata
ISOHDFS16&1.5&14\\
ISOHDFS23&1.2&65\\
ISOHDFS25&3.5&52\\
ISOHDFS27&3.6&42\\ 
ISOHDFS28&0.8&20\\
ISOHDFS38&18&600\\
ISOHDFS39&52&160\\
ISOHDFS53&7.0&40\\ 
ISOHDFS55&5.0&33\\
ISOHDFS60&18.5&63\\
ISOHDFS62&4.8&31\\
\enddata
\end{deluxetable}

\begin{figure}[hbtp]
\figurenum{1}
\plotone{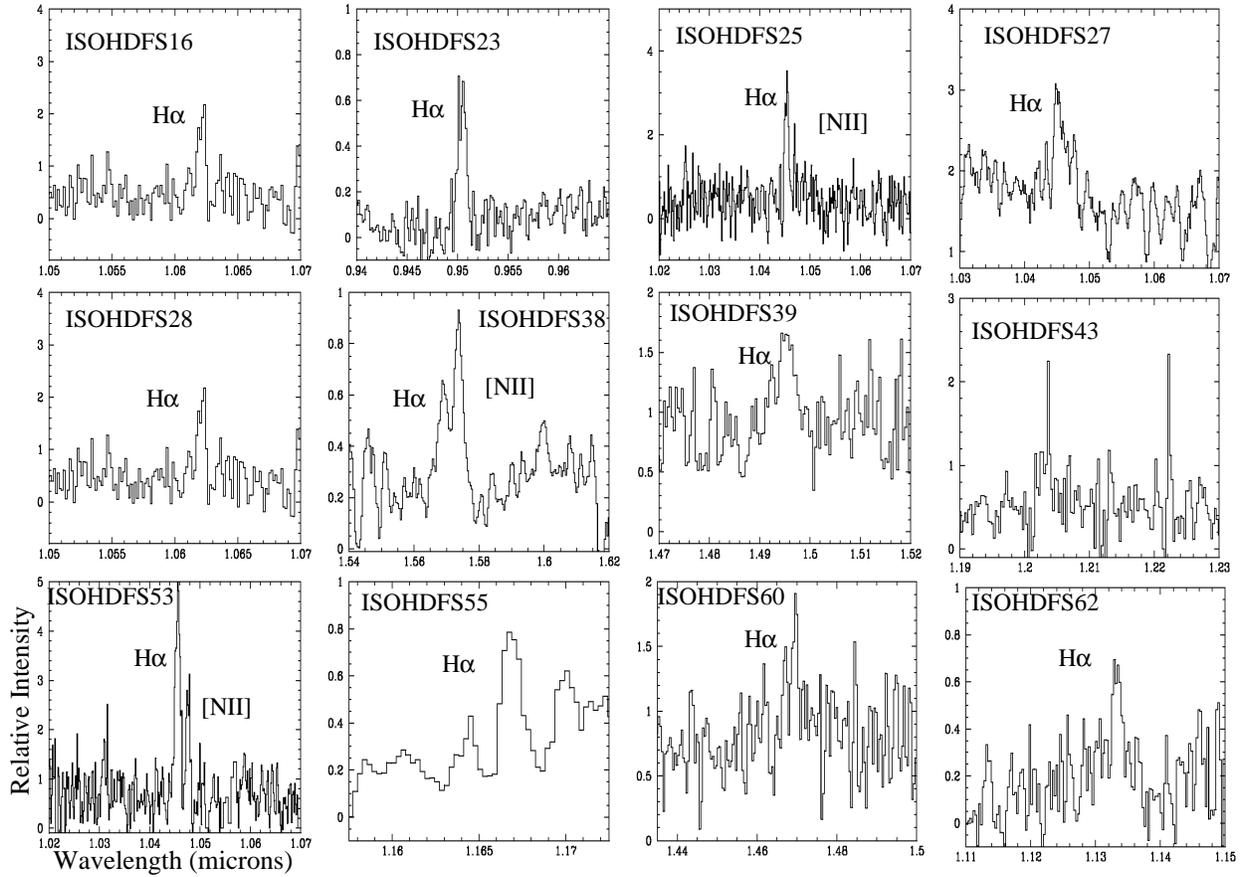}
\caption{ISAAC-VLT spectra. The H$_{\alpha}$ and [NII] (wherever resolved)
lines are indicated.}
\end{figure}

\begin{figure}[hbtp]
\figurenum{2}
\plotone{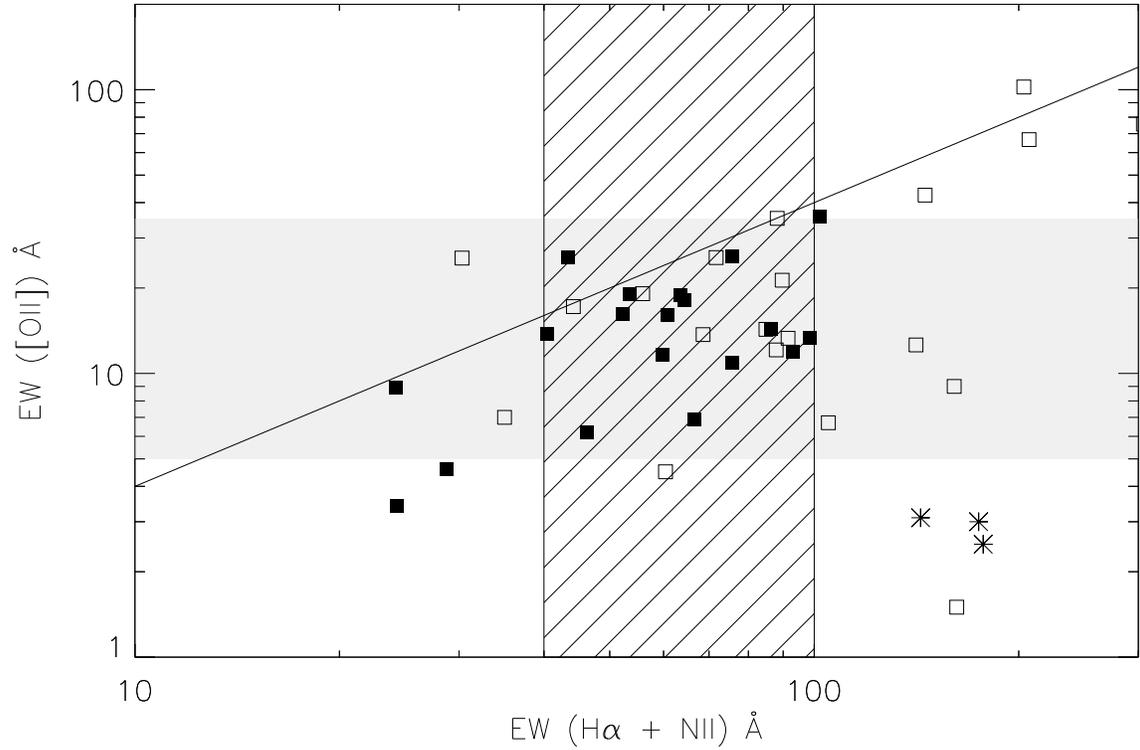}
\caption{EW(OII) vs. EW(H$_{\alpha} +$ NII) diagram (from Poggianti and
Wu (2000), filled squares e(a) galaxies, open squares non-e(a) galaxies,
stars Seyferts). The black line corresponds to EW(OII)= 0.4 EW(Ha$+$NII)
found for nearby field galaxies by Kennicutt 1992.
The shaded bars represent the area of the VLT ISOHDFS
galaxies. The intersection of the two bars is the location of our
galaxies, indistiguishable from the dusty luminous e(a) galaxies.}
\end{figure}

\end{document}